\documentclass[final,authoryear,3p,times,10pt]{elsarticle}

\usepackage{multirow,setspace,times,amssymb,amsmath,graphicx,color,rotating,subfigure,url}
\usepackage{amsthm}
\usepackage{mathrsfs}
\usepackage{lineno,color}
\usepackage{natbib}
\usepackage{booktabs}%
\usepackage{longtable}%
\usepackage[table]{xcolor}
\usepackage{tabularx}
\usepackage{graphicx} %use graph format
\usepackage{epstopdf}
\usepackage{mathrsfs}
\usepackage{makecell}
\usepackage{bm}
\usepackage[bookmarks=true,colorlinks,linkcolor=blue,anchorcolor=blue,citecolor=blue,unicode]{hyperref}
\usepackage{bookmark}
\PassOptionsToPackage{unicode}{hyperref}
\PassOptionsToPackage{naturalnames}{hyperref}
\usepackage{threeparttable}

\newtheorem{proposition}{Proposition}

\hypersetup{CJKbookmarks=true}%

\begin{document}
	
\begin{frontmatter}
	%\newpage
	\title{Hierarchical contagions in the interdependent financial network
}
	
	\author[Add1,Add2]{William A. Barnett}
	\author[Add3,Add4]{Xue Wang}
	\author[Add5]{Hai-Chuan Xu\corref{TBL}}
	\ead{hcxu@ecust.edu.cn}
	\cortext[TBL]{Corresponding author at: School of Business, East China University of Science and Technology, 130 Meilong Road, Shanghai 200237, China}
	\author[Add5]{Wei-Xing Zhou}
	
	\address[Add1]{Department of Economics, University of Kansas, Lawrence, USA}
	\address[Add2]{Center for Financial Stability, New York, USA}
	\address[Add3]{Institute of Chinese Financial Studies, Southwestern University of Finance and Economics, Chengdu, China}
	\address[Add4]{Department of Economics, Emory University, Atlanta, USA}
	\address[Add5]{Department of Finance and Research Center for Econophysics, East China University of Science and Technology, Shanghai, China}

\begin{abstract}
We derive the default cascade model and the fire-sale spillover model in a unified interdependent framework. The interactions among banks include not only direct cross-holding, but also indirect dependency by holding mutual assets outside the banking system. Using data extracted from the European Banking Authority, we present the interdependency network composed of 48 banks and 21 asset classes. For the robustness, we employ three methods, called $\textit{Anan}$, $\textit{Ha\l{}a}$ and $\textit{Maxe}$, to reconstruct the asset/liability cross-holding network. Then we combine the external portfolio holdings of each bank to compute the interdependency matrix. The interdependency network is much denser than the direct cross-holding network, showing the complex latent interaction among banks. Finally, we perform macroprudential stress tests for the European banking system, using the adverse scenario in EBA stress test as the initial shock. For different reconstructed networks, we illustrate the hierarchical cascades and show that the failure hierarchies are roughly the same except for a few banks, reflecting the overlapping portfolio holding accounts for the majority of defaults. We also calculate systemic vulnerability and individual vulnerability, which provide important information for supervision and relevant management actions.
\end{abstract}


\end{frontmatter}

%\tableofcontents
\newpage

\section{Introduction}

In recent years, network models, systemic stress testing and financial stability have attracted growing interest both among scholars and practitioners \citep{FJ-Battiston-MartinezJaramillo-2018-JFinancStab}. Regular stress tests conducted by authorities, such as the European Banking Authority, aim to evaluate the performance of individual banks in adverse scenarios, which are microprudential. Macroprudential outcomes are not simply the summation of micropridential changes. For example, when financial innovation reduces the cost of diversification, this may trigger a transition from stationary return dynamic to a nonstationary one \citep{FJ-Corsi-Marmi-Lillo-2016-OperRes}. Therefore, to be truly macroprudential, it is necessary to assess the role of network contagion in potentially amplifying systemic risk \citep{FJ-Gai-Kapad-2019-OxfRevEconPolicy}.

\begin{figure}[!htb]
 	\centering
 	\includegraphics[width=0.9\textwidth]{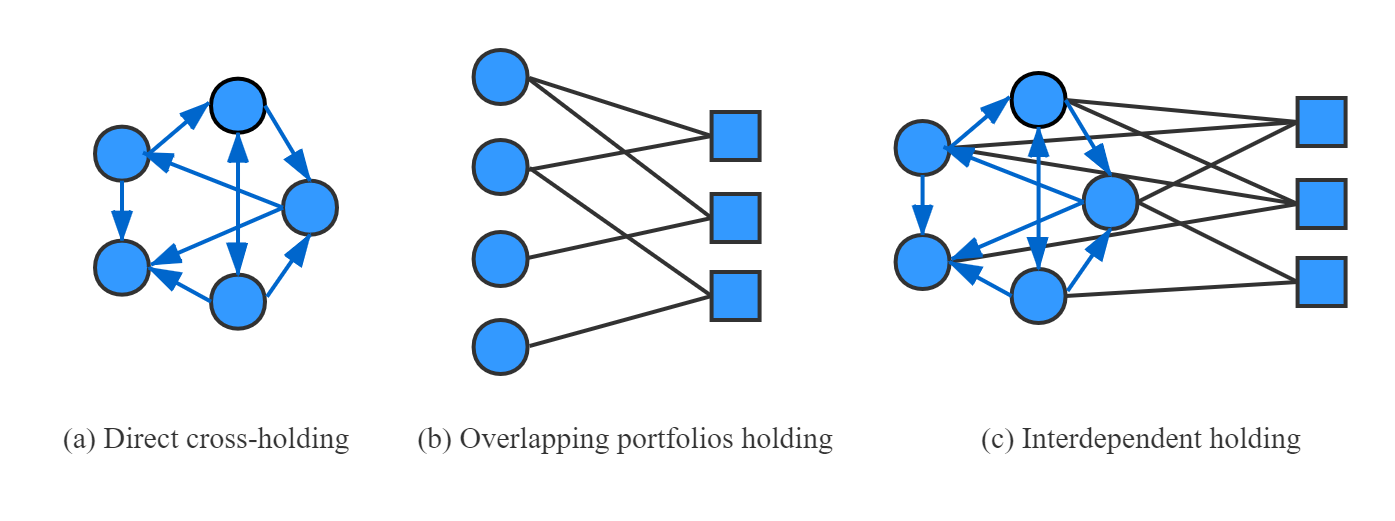}
 	\vspace{-8mm}
 	\caption{Illustrative examples showing 3 types of financial networks. Circles represent banks and squares stand for assets. } \label{Fig:Illustration}
 \end{figure}

There are different interactive channels among financial institutions. Figure~\ref{Fig:Illustration} illustrates three types of financial networks: (a) interbank network, (b) bank-asset bipartite network, and (c) interdependent network. The interbank network characterizes direct credit exposures to other banks and risk contagion can be caused by direct cross-holding. For example, \citet{FJ-Dungey-Flavin-LagoaVarela-2020-JBankFinanc} empirically analyze the transmission of shocks among global banks, domestic banks and the non-financial sector for 11 Eurozone countries. Apart from direct connection, it's apparent that banks are indirectly connected by holding overlapping portfolio outside the banking system as in Figure~\ref{Fig:Illustration} (b).  \citet{FJ-Barucca-Mahmood-Silvestri-2021-JFinancStab} empirically find significant overlapping equity and debt portfolios between different types of financial institution, providing evidence for the existence of a price-mediated channel of contagion between banks. The third type of network is much more complex, including not only direct cross-holding, but also indirect dependency by holding mutual assets as in Figure~\ref{Fig:Illustration} (c). This interdependency has been shown as a realistic source of uncertainty in systemic risk \citep{FJ-Roukny-Battiston-Stiglitz-2018-JFinancStab}. Furthermore, \citet{FJ-Elliott-Golub-Jackson-2014-AmEconRev} study cascading failures in an equilibrium model of interdependent financial network.
 
Since interdependent network model provides two contagion channels and is more realistic, it is worthy of further study.  \cite{FJ-Elliott-Golub-Jackson-2014-AmEconRev} propose a default cascade model, where an initial shock causes at least one bank failed, then the bankruptcy loss of this failure spreads through the interbank lending network. When the equity values of connected banks fall below certain critical threshold, the cascade will occur. Unlike losses caused by price fluctuations, bankruptcy losses are discontinuous. Therefore, \cite{FJ-Elliott-Golub-Jackson-2014-AmEconRev} model the cascade in a discontinuous way. However, even in the absence of bank failures, risk can still spread through fire sales in a continuous way (see \cite{FJ-Cifuentes-Ferrucci-Shin-2005-JEurEconAssoc}, \cite{FJ-Nier-Yang-Yorulmazer-Alentorn-2007-JEconDynControl}; \cite{FJ-Allen-Babus-2009-book}, \cite{FJ-Caccioli-Shrestha-Moore-Farmer-2014-JBankFinanc}, \cite{FJ-Greenwood-Landier-Thesmar-2015-JFinancEcon}, \cite{FJ-Glasserman-Young-2016-JEconLit}, \cite{FJ-Cont-Schaanning-2019-JBankFinanc}, and \cite{FJ-Duarte-Eisenbach-2021-JFinanc}). In these models, a large negative shock leads to an increase in leverage. To reduce their leverages, banks sell assets to pay off debts. These asset fire sales have a price impact, hence cause a fire-sale spillover. Although the fire-sale spillover has been studied well, none of them model the contagion in an interdependent network. For example, some construct an interbank network and model the fire-sale effect on an aggregate external asset (e.g., \cite{FJ-Cifuentes-Ferrucci-Shin-2005-JEurEconAssoc}, \cite{FJ-Nier-Yang-Yorulmazer-Alentorn-2007-JEconDynControl}, \cite{FJ-Allen-Babus-2009-book} and \cite{FJ-Glasserman-Young-2016-JEconLit}). Others model financial contagion from overlapping portfolios but without interbank interaction (e.g., \cite{FJ-Caccioli-Shrestha-Moore-Farmer-2014-JBankFinanc}, \cite{FJ-Greenwood-Landier-Thesmar-2015-JFinancEcon}, \cite{FJ-Cont-Schaanning-2019-JBankFinanc} and \cite{FJ-Duarte-Eisenbach-2021-JFinanc}). In this paper, we derive the default cascade model and the fire-sale spillover model in a unified interdependent framework. We prove that these two loss mechanisms are essentially consistent, although one propagates losses in a discontinuous way and the other in a continuous way. 

In addition, we slightly revised the interdependent framework of \citet{FJ-Elliott-Golub-Jackson-2014-AmEconRev} by separating the bank's ``value'' delivered to final investors outside the banking system to external liabilities and equity value. Such division is in line with the balance sheet and can make clear the bank equity value in the general sense, although it does not change the derived form of interdependency matrix. This modification also facilitates empirical research for the European banking system, because the European Banking Authority (EBA) dataset does not provide liability items, but only asset items and some equity items such as Tier 1 capital. 

Then, we empirically identify cascade hierarchies in an interdependent financial network. We integrate microprudential stress test and macroprudential stress test together for the European banking system. Considering that the EBA's stress test is microprudential for individual banks, we perform macroprudential stress test by using the adverse scenario in EBA's stress test as the initial shock. Since granular data on interbank credit exposures is not public, we employ 3 reconstruction methods to form the cross-holding network and then study contagion hierarchies comparatively. 

The remainder of the paper is organized as follows. Section~\ref{S:Lit} presents the literature review. Section~\ref{S:Meth} introduces the model and method of identifying cascade hierarchies. Section~\ref{S:Emp} shows the data and the empirical analyses. Section~\ref{S:Conclude} concludes the paper.

\section{Literature review}
\label{S:Lit}
As in Figure~\ref{Fig:Illustration}, we review existing literature about network contagion according to the network structures adopted.  

\subsection{Interbank network contagions}

This kind of model shows that contagion can be caused by direct credit exposures among banks. \citet{FJ-Rogers-Veraart-2013-ManageSci} model financial market as a directed graph of interbank obligations and study the occurrence of systemic risk. \citet{FJ-Gai-Kapadia-2010-ProcRSocA-MathPhysEngSci} develop an analytical network contagion model and suggest that financial systems exhibit a robust-yet-fragile tendency. That is, while the probability of contagion may be low, the influences can be extremely widespread when problems occur. Similarly, \citet{FJ-Acemoglu-Ozdaglar-TahbazSalehi-2015-AmEconRev} argue that the extent of financial contagion exhibits a form of phase transition. In addition, many studies focus on how interbank network topology creates instability \citep{FJ-Bardoscia-Battiston-Caccioli-Caldarelli-2017-NatCommun, FJ-Eboli-2019-JEconDynControl}. \citet{FJ-Zhang-Fu-Lu-Wang-Zhang-2021-JFinancStab} find that network connectedness of banks strengthens the relationship between liquidity creation and systemic risk. \citet{FJ-Brunetti-Harris-Mankad-Michailidis-2019-JFinancEcon} study the interbank market around the 2008 financial crisis and find that the correlation network and the physical credit network behavior different. During the crisis, the correlation network displays an increase in connection, while the physical credit network shows a marked decrease in connection.

\subsection{Overlapping portfolio contagions}

When a bank suffers a negative shock to its equity, a natural way to return to target leverage is to sell assets. \citet{FJ-Greenwood-Landier-Thesmar-2015-JFinancEcon} present a model in which fire sales propagate shocks across banks. \citet{FJ-Huang-Vodenska-Havlin-Stanley-2013-SciRep} build a bipartite banking network model composed of banks and assets and present a cascading failure describing the risk propagation process during crises. Similarly, \citet{FJ-Caccioli-Shrestha-Moore-Farmer-2014-JBankFinanc} show the amplification of financial contagion due to the combination of overlapping portfolios and leverage, in terms of a generalized branching process. Furthermore, for quantifying the potential exposure to indirect contagion arising from deleveraging of assets in stress scenarios, \citet{FJ-Cont-Schaanning-2019-JBankFinanc} propose two indicators. \citet{FJ-Vodenska-Aoyama-Becker-Fujiwara-Iyetomi-Lungu-2021-JFinancStab} build a bipartite network with weighted links between banks and assets based on sovereign debt holdings, and then model the systemic risk propagation. 

\subsection{Interdependent network contagions}

This kind of model investigates how these two channels (the interbank channel and the overlapping channel) propagate individual defaults to systemic cascading failures. \citet{FJ-Caccioli-Farmer-Foti-Rockmore-2015-JEconDynControl} argue that neither channel of contagion results in large effects on its own. In contrast, when both channels are active, defaults are much more common and have large systemic effects. \citet{FJ-Aldasoro-DelliGatti-Faia-2017-JEconBehavOrgan} likewise suggest that contagion occurs through deleveraging and interbank connection. The interdependent network models are also applied to characterize contagions in reinsurance and derivatives markets \citep{FJ-KlagesMundt-Minca-2020-ManageSci, FJ-Paddrik-Rajan-Young-2020-ManageSci}. 

\citet{FJ-Elliott-Golub-Jackson-2014-AmEconRev} study cascading failures in an interdependent financial network. They show that discontinuous changes in asset values trigger further failures. Furthermore, when banks face potentially correlated risks from outside the financial system, the interbank connections can share these risks, but they also create the channels by which shocks can be propagated \citep{FJ-Elliott-Georg-Hazell-2021-JEconTheory}. \citet{FJ-Poledna-MartinezJaramillo-Caccioli-Thurner-2021-JFinancStab} model the Mexican banking system as a multi-layer network of direct interbank exposures and indirect overlapping exposures, and estimate the mutual influence of different channels of contagion. In addition, some studies find that the overlapping portfolio holding by banks accounts for the majority of defaults. \citet{FJ-Chen-Liu-Yao-2016-OperRes} confirm that the market liquidity effect has a great potential to cause systemic contagion. \citet{FJ-Dungey-Flavin-LagoaVarela-2020-JBankFinanc} show that deleveraging speed and concentration of illiquid assets play a critical role in cascades. \citet{FJ-Ma-Zhu-Wu-2021-QuantFinanc}  further prove that illiquidity is a critical factor in triggering risk contagion and that higher interbank leverage can cause larger losses for both the banks and the external assets. Our results are consistent with these literature, in the sense that the general contagion hierarchies are mainly determined by the overlapping channel, while the structure of interbank network is also important for some specific banks.

\section{Models}
\label{S:Meth}

\subsection{The default cascade model}

The model follows \citet{FJ-Elliott-Golub-Jackson-2014-AmEconRev}, but separates the ``value'' in their paper, that any bank delivers to final investors outside the system of cross-holding, to external liabilities and equity value. Concretely, for every bank, its assets are divided into external assets and interbank assets, and its liabilities are divided into external liabilities and interbank liabilities. The equity value is the difference between its total assets and its total liabilities. Table~\ref{Tb:Balance:Sheet} illustrates a balance sheet based on this. 

\begin{table}[!htb]
	\centering
	\caption{Balance Sheet of Bank $i$.}
	\smallskip
	\begin{tabular}{lclc}
			\toprule
			Assets               &     &   Liabilities             & \\ \midrule
			\rule{0pt}{15pt} External assets  &  $\sum_{k} D_{i k} p_{k} $ & External liabilities   & $l_i^{(e)} V_{i}$ \\ 
			\rule{0pt}{15pt} Interbank assets&  $a_{i}\equiv\sum_{j} C_{i j} V_{j}$   & Interbank liabilities &  $l_{i}\equiv\sum_{j} C_{j i} V_{i}$ \\
			\rule{0pt}{15pt}                          &      & Net worth               &  $v_{i}$ \\
			\bottomrule
	\end{tabular}
	%\tabnote{\textsuperscript{a}This footnote shows how to include footnotes to a table if required.}
	\label{Tb:Balance:Sheet}
\end{table}		

 Assume that there are $N$ banks and $M$ external assets. The current value of asset $k$ in the system is denoted $p_k$. Let $D_{ik}\geq0$ be the fraction of the value of asset $k$ held directly by bank $i$ and let $\mathbf{D}$ denote the matrix whose entry is equal to  $D_{ik}$. A bank can also hold shares of other banks. Let $C_{ij}\geq0$ be the fraction of bank $j$ owned by bank $i$, where $C_{ii}=0$ for each $i$. The cross-holding matrix $\mathbf{C}$ can be viewed as a network in which there is a directed link from $j$ to $i$ if cash flows in that direction, in other words, if $i$ owns a positive share of $j$. 
 
 Let $V_i$ be the total asset value of bank $i$. This is equal to the value of external assets holding by bank $i$ plus the value of its claims on other banks:
\begin{equation}\label{Eq:V1}
	V_{i}=\sum_{k} D_{i k} p_{k}+\sum_{j} C_{i j} V_{j}.
\end{equation}

In matrix notation, Equation~(\ref{Eq:V1}) can be written as
\begin{equation}\label{Eq:V2}
	\mathbf{V}=\mathbf{D} \mathbf{p}+\mathbf{C V}
\end{equation}
and solved to yield 
\begin{equation}\label{Eq:V3}
	\mathbf{V}=(\mathbf{I}-\mathbf{C})^{-1} \mathbf{D} \mathbf{p},
\end{equation}
where $\mathbf{I}$ is an identity matrix.

On the other hands, the total value of bank $i$ is also equal to its total liabilities plus its equity value $v_i$. Its total liabilities constitute of interbank liabilities $\sum_{j} C_{j i} V_{i}$ and external liabilities $l_i^{(e)} V_{i}$, where $l_i^{(e)}$ is the ratio of external liabilities to total assets. Hence, the equity value of bank $i$:
\begin{equation}\label{Eq:v1}
	v_{i}=\sum_{j} C_{i j} V_{j}-\sum_{j} C_{j i} V_{i}+\sum_{k} D_{i k} p_{k} -  l_i^{(e)} V_{i}.
\end{equation}

Now we denote the capital ratio (i.e. ratio of equity value to total value) of bank $i$ as  $\widehat{C}_{ii}$, then
\begin{equation}\label{Eq:C:hat}
	\widehat{C}_{ii} \equiv 1 - l_i^{(e)} - \sum_{j \in N} C_{j i}.
\end{equation}
Let $\widehat{\mathbf{C}}$ be the diagonal matrix of capital ratios, where each diagonal entry is $\widehat{C}_{ii}$. Hence, Equation~(\ref{Eq:v1}) can be written in matrix notation as 
\begin{equation}\label{Eq:v2}
	\mathbf{v}=\mathbf{C V}-(\mathbf{I}-\widehat{\mathbf{C}}) \mathbf{V}+\mathbf{D} \mathbf{p}=(\mathbf{C}-(\mathbf{I}-\widehat{\mathbf{C}})) \mathbf{V}+\mathbf{D} \mathbf{p}.
\end{equation}
Substituting for the total asset value V from (\ref{Eq:V3}), this becomes
\begin{equation}\label{Eq:v3}
	\begin{aligned}
		\mathbf{v} &=(\mathbf{C}-\mathbf{I}+\widehat{\mathbf{C}})(\mathbf{I}-\mathbf{C})^{-1} \mathbf{D} \mathbf{p}+\mathbf{D} \mathbf{p}=(\mathbf{C}-\mathbf{I}+\widehat{\mathbf{C}}+(\mathbf{I}-\mathbf{C}))(\mathbf{I}-\mathbf{C})^{-1} \mathbf{D} \mathbf{p}\\
		&=\widehat{\mathbf{C}}(\mathbf{I}-\mathbf{C})^{-1} \mathbf{D} \mathbf{p}=\mathbf{A D p}.
	\end{aligned}	
\end{equation}
Here we refer to $\mathbf{A}=\widehat{\mathbf{C}}(\mathbf{I}-\mathbf{C})^{-1}$ as the interdependency matrix.

As in \citet{FJ-Elliott-Golub-Jackson-2014-AmEconRev}, banks will lose some value in discontinuous ways if their values fall below certain critical thresholds. In fact, it's these discontinuities that lead to cascading failures. If the equity value $v_i$ of a bank $i$ falls below some threshold level $\underline{v}_{i}$, then the bank is said to fail and incurs failure costs $\beta_{i}\underline{v}_{i}$ with $\beta \in [0,1]$. In many situations, a natural cap for $\beta_i$ is 1. That is, the maximum failure cost for bank $i$ will not exceed its equity value at the time of failure. 

The valuations in (\ref{Eq:V3}) and (\ref{Eq:v3}) are similar when we include the discontinuous failure costs, and so the total value of bank $i$ becomes 
\begin{equation}\label{Eq:V1:cost}
	V_{i}=\sum_{j \neq i} C_{i j} V_{j}+\sum_{k} D_{i k} p_{k}-\beta_{i}\underline{v}_{i} I_{v_{i}<\underline{v}_{i}},
\end{equation}
where $I_{v_{i}<\underline{v}_{i}}$ is an indicator variable taking value 1 if $v_{i}<\underline{v}_{i}$ and value 0 otherwise.

Then a failure version of (\ref{Eq:V3}) is:
\begin{equation}\label{Eq:V2:cost}
	\mathbf{V}=(\mathbf{I}-\mathbf{C})^{-1}(\mathbf{D} \mathbf{p}-\mathbf{b}(\mathbf{v})),
\end{equation}
where $ b_i(\mathbf{v})= \beta_{i}\underline{v}_{i} I_{v_{i}<\underline{v}_{i}}$. Correspondingly, (\ref{Eq:v3}) is re-expressed as
\begin{equation}\label{Eq:v1:cost}
	\mathbf{v}=\widehat{\mathbf{C}}(\mathbf{I}-\mathbf{C})^{-1}(\mathbf{D} \mathbf{p}-\mathbf{b}(\mathbf{v}))=\mathbf{A}(\mathbf{D} \mathbf{p}-\mathbf{b}(\mathbf{v})).
\end{equation}
An entry $A_{ij}$ of the interdependency matrix describes the proportion of $j$'s costs that $i$ pays when $j$ fails as well as $i$'s claims on the external assets that $j$ directly holds.

This model differs from those studied in \cite{FJ-Eisenberg-Noe-2001-ManageSci} and \cite{FJ-Rogers-Veraart-2013-ManageSci}. In their models, every bank pays the minimum of what it owes and what it has, such that a recursive repayment vector (also called clearing vector) is obtained. Differently, in the framework of \cite{FJ-Elliott-Golub-Jackson-2014-AmEconRev}, one bank's value decreases by a certain amount arising from other interdependent banks' failure costs, such that a recursive equity value vector is obtained as (\ref{Eq:v1:cost}). Using Tarski's fixpoint theorem \citep{FJ-Tarski-1955-PJM}, we can prove the existence of fixed points for Equation~(\ref{Eq:v1:cost}). 

We define an operator $T: \mathbb{R}^{N} \rightarrow \mathbb{R}^{N}$ as
\begin{equation}\label{Eq:T}
	T\mathbf{v}=\mathbf{A}(\mathbf{D}\mathbf{p}-\mathbf{b}(\mathbf{v})).
\end{equation} 
We also set $\mathbf{g}:=\mathbf{A}(\mathbf{D}\mathbf{p}-\beta\underline{\mathbf{v}})$ and $\mathbf{v}_0 := \mathbf{A}\mathbf{D}\mathbf{p}$, then an interval is defined as $H:=[\mathbf{g}, \mathbf{v}_0]$. Here $\mathbf{v}_0$ indicates the initial equity values of banks. Note that $\mathbf{p}$, $\underline{\mathbf{v}}$, $\mathbf{g}$, and $\mathbf{v}_0$ are all given and unchanged in the absence of fire-sales.
\begin{proposition}\label{Pro:existence}
	$T$ is a self-map on $H$ with least fixed point $\mathbf{v_{*}}$ and greatest fixed point $\mathbf{v^{*}}$ in $H$. In addition, 
	\item (i) the sequence $\{T^{k}\mathbf{g}\}$ converges up to $\mathbf{v_{*}}$ in a finite number of iterations and 
	\item (ii) the sequence $\{T^{k}\mathbf{v}_0\}$ converges down to $\mathbf{v^{*}}$ in a finite number of iterations.
\end{proposition}

The proof can be found in \ref{S:Proof}. As stated in \cite{FJ-Elliott-Golub-Jackson-2014-AmEconRev}, the map~(\ref{Eq:T}) can exist multiple equilibria due to discontinuous failure costs. What we focus on is the best-case equilibrium in terms of having the maximum value $\mathbf{v^{*}}$. Proposition~\ref{Pro:existence} (ii) now guarantees to find the best-case equilibrium using the following algorithm.

\subsection{Identifying cascade hierarchies}

Based on the interdependent model, we can trace the propagation path initiated by a specific shock. At step $t$, let $Z_t$ be the set of failed banks. Initialize $Z_0 = \emptyset$ and $\mathbf{\underline{v}}=\theta\mathbf{v}_{0}$. Assume an adverse scenario that causes prices of mutual assets to decline. Then the cascade hierarchies can be identified as following. At each step $t\geq1$:
\begin{enumerate}
  \item Let $\mathbf{\widetilde{b}}_{t-1}$ be a vector with element $\widetilde{b}_{i} = \beta_{i}\underline{v}_{i}$ if $i\in Z_{t-1}$ and 0 otherwise.
  \item If the entries of the following vector is negative, then the corresponding banks are added to $Z_t$.
  \begin{equation}\label{Eq:Z}
  	\mathbf{A}\left[\mathbf{D} \mathbf{p}-\widetilde{\mathbf{b}}_{t-1}\right]-\underline{\mathbf{v}}.
  \end{equation}
  \item Terminate if $Z_t = Z_{t-1}$, otherwise return to step 1.
\end{enumerate}
When this algorithm terminates at step $T$, the sets $Z_1, Z_2,...,Z_T$ correspond to the failure hierarchies in the best-case
equilibrium.

\subsection{Link to the fire-sale spillover model}

The model above describes a kind of risk contagion path by default cascade. Specifically, banks will default and lose value in a discontinuous way when their equity values fall below certain critical thresholds. It is the discontinuity that leads to cascading failures. However, even in the absence of bank failures, risk can still spread through fire sales (see \cite{FJ-Cifuentes-Ferrucci-Shin-2005-JEurEconAssoc}, \cite{FJ-Nier-Yang-Yorulmazer-Alentorn-2007-JEconDynControl}; \cite{FJ-Allen-Babus-2009-book}, \cite{FJ-Caccioli-Shrestha-Moore-Farmer-2014-JBankFinanc}, \cite{FJ-Greenwood-Landier-Thesmar-2015-JFinancEcon}, \cite{FJ-Glasserman-Young-2016-JEconLit}, \cite{FJ-Cont-Schaanning-2019-JBankFinanc}, and \cite{FJ-Duarte-Eisenbach-2021-JFinanc}). In these models, banks are assumed to hold a target leverage ratio (or a regulatory minimum capital ratio). When a bank violates this constraint due to a negative shock, it sells assets to pay off partial debts so as to reduce its leverage. This will cause a fire-sale spillover. Now we will show that this fire-sale contagion can also be modeled in the same framework as above. 

We introduce some additional notations. Let $W_{ik}\geq0$ be the proportion of asset $k$ in all assets held by bank $i$ and let $\mathbf{W}$ denote the matrix whose entry is equal to $W_{ik}$. Note that the matrix $\mathbf{W}$ is defined relative to total assets of bank $i$ such that $\sum_{k} W_{ik}=1$, while the matrix $\mathbf{D}$ is defined relative to all banks such that $\sum_{i} D_{ik}=1$. There are two parts of losses emerging in the fire-sale scenario, i.e., direct loss caused by initial shock and deleveraging loss. First, an initial exogenous shock $\mathbf{r_1}$ leads to a loss rate for bank external assets as
\begin{equation}\label{Eq:b1}
	\mathbf{b_1} = \mathbf{Wr_1},
\end{equation}
where $\mathbf{r_1}$ is an $M \times 1$ vector indicating decline ratios of asset values. When the shock is negative, the dollar value of bank assets will decrease by \footnote{Instead of using loss rate $\mathbf{b_1}$, this direct loss can also be calculated in terms of the changes of asset values, that is, $\mathbf{s_1} = \mathbf{D}(\mathbf{p_0} - \mathbf{p_1})$. However, it's not convenient to describe the assumption for deleveraging as follows. }
\begin{equation}\label{Eq:s1}
	\mathbf{s_1} = \mathbf{Dp_0}\circ \mathbf{b_1}, 
\end{equation}
where $\circ$ refers to the Hadamard product (i.e. component-wise multiplication). This is the direct loss. Then, banks sell assets to return to their target leverages. Recall that the diagonal matrix $\widehat{\mathbf{C}}$ denotes capital ratios of equity to total assets. Then the diagonal matrix of leverage ratios of debt to equity is denoted by $\mathbf{\Lambda}=\widehat{\mathbf{C}}^{-1} - \mathbf{I}$, where $\mathbf{I}$ is an identity matrix. To deleverage, banks have to sell assets and pay off debts with amount\footnote{For simplicity, we assume that only external assets will be sold. If some elements of $\mathbf{\pi}$ exceed the current value of external assets, the target leverage will not be fully returned. In this case, the bank will liquidate all of its external assets. To take this situation into account, the dollar amount of assets should be modified as $\mathbf{\pi} = \min (\mathbf{\Lambda s_1}, \mathbf{Dp_0}\circ (\mathbf{1}-\mathbf{b_1}))$}
\begin{equation}\label{Eq:pi}
	 \mathbf{\pi} = \mathbf{\Lambda s_1}.
\end{equation}
The intuition of this formula is straightforward. Suppose a bank with equity of 1 and debt of 9 losses 0.5 due to a shock, bring its equity to 0.5. The bank has to sell 4.5 assets and use the cash to pay off 4.5 debts to the prior leverage of 9. Now we must decide which assets are sold. As in \cite{FJ-Greenwood-Landier-Thesmar-2015-JFinancEcon}, we assume that banks sell assets proportionately to their current holdings, which means that the matrix $\mathbf{W}$ remains constant. Let $\mathbf{\phi}$ be the $M \times 1$ vectors of dollar amount of asset sales by all banks, then 
\begin{equation}\label{Eq:phi}
	\mathbf{\phi} = \mathbf{W^{\prime}\pi}.
\end{equation}
Now we further assume that these asset sales will generate price impact according to a linear model as in \cite{FJ-Greenwood-Landier-Thesmar-2015-JFinancEcon},
\begin{equation}\label{Eq:r2}
	\mathbf{r_2} = \mathbf{\Omega\phi},
\end{equation}
where $\mathbf{\Omega}$ is a diagonal matrix of price impact ratios. Under the shock $\mathbf{r_2}$, we can calculate the loss rate in this deleveraging round as 
\begin{equation}\label{Eq:b2}
	\mathbf{b_2} =  \mathbf{Wr_2},
\end{equation}
or re-written as
\begin{equation}\label{Eq:b2:2}
	\mathbf{b_2} = \mathbf{W\Omega W^{\prime}\Lambda Dp_0}\circ \mathbf{b_1}.
\end{equation}
One can iterate multiple rounds of deleverage to get loss rates $\mathbf{b_3},\dots, \mathbf{b_{\infty}}$. In the limit, combining with (\ref{Eq:v3}), we obtain the equity value 
\begin{equation}\label{Eq:v:fire}
	\mathbf{v} = \mathbf{ADp_0}\circ (\mathbf{1}- \mathbf{b_1}) \circ (\mathbf{1}- \mathbf{b_2}) \circ \dots \circ (\mathbf{1}- \mathbf{b_{\infty}}),
\end{equation}
where $\mathbf{1}$ is a vector of ones. 

Now we have derived the default cascade model and the fire-sale spillover model in a unified interdependent framework. Compared (\ref{Eq:v:fire}) with (\ref{Eq:v1:cost}), we prove that these two loss mechanisms are essentially consistent. In the default cascade model, an initial shock causes at least one bank failed, bring the system loss to $\mathbf{Ab}(\mathbf{v})$. Then the cascade is activated if the equity values of the interconnected banks fall below some threshold level as in (\ref{Eq:Z}). Similarly, in the fire-sale spillover model, an initial shock causes the system loss $\mathbf{ADp_0}\circ \mathbf{b_1}$. Then the second round loss  $\mathbf{ADp_0}\circ (\mathbf{1}- \mathbf{b_1}) \circ \mathbf{b_2}$ occurs through deleveraging. Although there is no default in the fire-sale spillover model, the mechanism of value loss is similar with the default model. In next section, we will present an empirical analysis based on the default cascade model.

\section{Empirical analyses for European banking system}
\label{S:Emp}
\subsection{Data}
We use data collected by the European Banking Authority (EBA) for the 2018 EU-wide stress test. This public dataset covers a sample of 48 banks in 15 countries in the European Union and European Economic Area at the highest level of consolidation. Table~\ref{Tb:Bank:list} lists 48 banks and countries they belong to. This dataset not only provides the actual balance sheet figures and their International Financial Reporting Standard (IFRS) 9 restated figures, but also covers a three-year horizon baseline and adverse scenarios, which take the end-2017 data as the starting point\footnote{https://www.eba.europa.eu/risk-analysis-and-data/eu-wide-stress-testing/2018}. 	

The actual and restated figures give the exposure values in various asset classes. Table~\ref{Tb:Asset:Class} lists 21 asset classes that we collect from the EBA dataset and provides corresponding EBA items and EBA exposure codes for each type of asset. Among them, type 2100 indicates the aggregated claims on other credit institutions that one bank holds. However, granular exposure data on banking networks is not public. The credit exposure networks can be reconstructed by some inference methods using only aggregated relational data \citep{FJ-Anand-vanLelyveld-Banai-Friedrich-Garratt-Halaj-Fique-Hansen-Jaramillo-Lee-MolinaBorboa-Nobili-Rajan-Salakhova-Silva-Silvestri-deSouza-2018-JFinancStab}.  The other 20 classes are the external assets mutually holding by 48 banks. 

The adverse scenario gives the corresponding exposure values of various asset classes under some assumed macroeconomic shocks, including a growth in gross domestic product (GDP) in the EU of -1.2\%, -2.2\% and 0.7\% as of 2018, 2019 and 2020 respectively. This adverse scenario can be viewed as a ideal initial shock with which the proposed hierarchical contagion model will be tested. 

\begin{table}[!htb]
	\centering
	\caption{Asset classes and their EBA data reference codes.}
	\smallskip
	\begin{threeparttable}
	\begin{tabular}{lll}
			\toprule
			 EBA Item  &	EBA Exposure &	Asset classes \\ \midrule
			\multirow{5}{*}{183203, 183303} & 1100	& Central banks and central governments \\
			& 1200	& Regional governments or local authorities \\
			& 1300	& Public sector entities \\
			& 1400	& Multilateral Development Banks \\
			& 1500	& International Organisations \\\midrule
			\multirow{3}{*}{183904, 183905} & 1700	& General governments \\
			& \textbf{2100}	& \textbf{Credit institutions} \\
			& 2200	& Other financial corporations\\\midrule
			\multirow{7}{*}{183203, 183303} & 3000	& Corporates (Credit Risk) / Non Financial corporations (NPE- Forbearance) \\
			& 4110	& Retail - Secured by real estate property - SME\\
			& 4120	& Retail - Secured by real estate property - Non SME\\
			& 4200	& Retail  - Qualifying Revolving\\
			& 4310	& Retail - Other - SME\\
			& 4320	& Retail - Other - Non SME\\
			& 4500	& Retail - SME\\\midrule
			183904, 183905	& 4700	& Households \\\midrule
			\multirow{5}{*}{183203, 183303} & 5000	& Secured by mortgages on immovable property \\
			& 6400	& Items associated with particularly high risk\\
			& 6500	& Covered bonds\\
			& 6600	& Claims on institutions and corporates with a ST credit assessment\\
			& 6700	& Collective investments undertakings (CIU)\\
			\bottomrule
	\end{tabular}
	\begin{tablenotes}
		\footnotesize
		\item[]Note: NPE means non-performing exposure; SME means small and medium-sized enterprises; ST means short-term. 
	\end{tablenotes}
	\end{threeparttable}
		\label{Tb:Asset:Class}
\end{table}		

\begin{figure}[!htb]
	\centering
	\includegraphics[width= 0.45\textwidth]{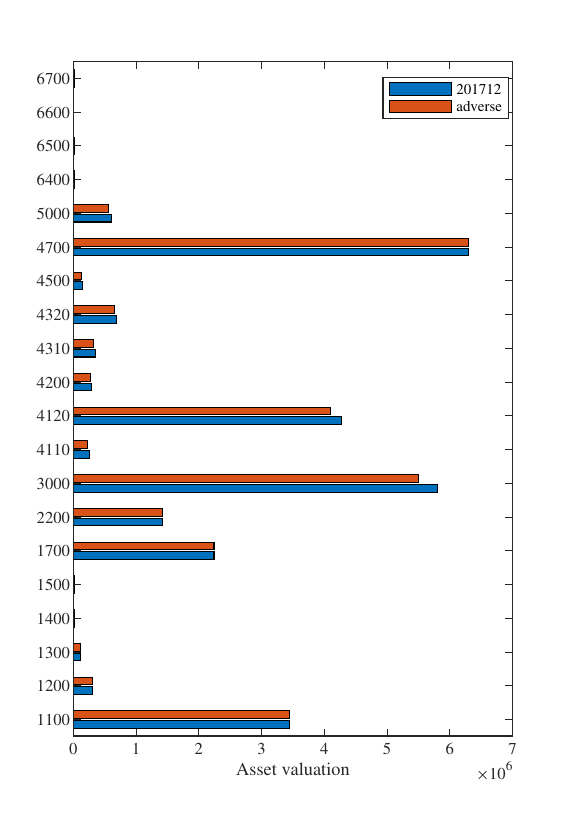}
	\includegraphics[width= 0.45\textwidth]{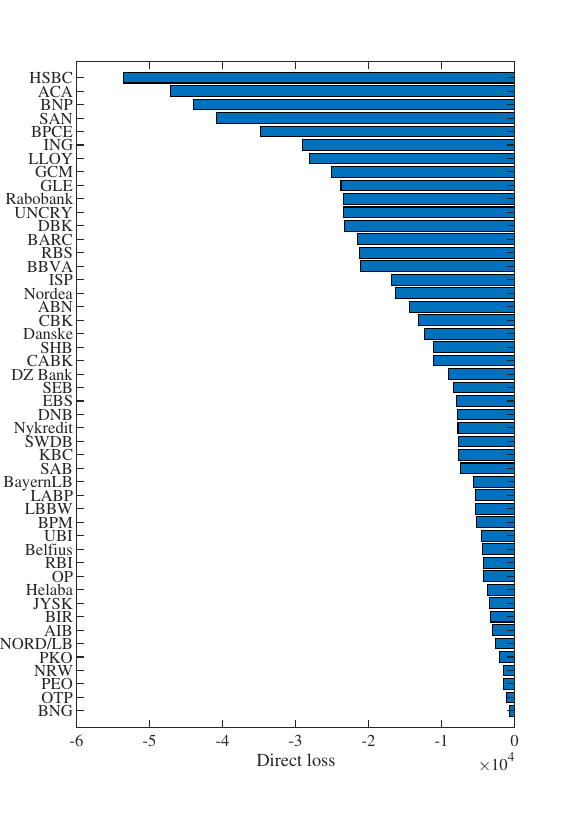}
	\caption{The asset values and the direct loss. Left: the values of 20 asset classes at the end of 2017 and at the end of 2020 in the adverse scenario test by EBA. Right: the direct loss for 48 banks in the adverse scenario.} \label{Fig:loss}
\end{figure}

Figure~\ref{Fig:loss} shows the values $\mathbf{p}$ of 20 external asset classes and the direct loss $\mathbf{D\Delta p}$ for 48 banks in the adverse scenario. The assets whose values drop the most are 3000 (Corporates / Non Financial corporations) and 4120 (Retail - Secured by real estate property - Non SME). It reflects that corporations are undergoing some degree of financial distress in the adverse scenario. The depreciation of these assets directly leads to equity losses of banks. It's reasonable that bigger banks will lose more, such as HSBC, ACA, BNP, and SAN (see right panel of  Figure~\ref{Fig:loss}).

Table~\ref{Tb:CET} reports additional key statistics based on balance sheet data from EBA. Under the adverse scenario, the risk-weighted Basel CET 1 capital ratio (equity over risk-weighted total assets) of the European banking system decreases from 14.42\% to 10.32\%, and the Basel leverage ratio (equity over unweighted total assets)\footnote{Note that the Basel leverage ratio is not defined as the well known financial leverage ratio.} decreases from 5.31\% to 4.37\%. 

\begin{table}[!htb]
	\centering
	\caption{The distribution of Basel CET 1 ratio and Basel leverage ratio for 48 banks at the end of 2017 and at the end of 2020 in the adverse scenario test by EBA.}
	\smallskip
	\begin{tabular}{lrrrrr}
		\toprule
		    & System & Min & Median & Mean & Max \\
		\midrule
		Basel CET 1 Ratio (2017) &  14.42 &  11.38 &  15.40 &  16.34 &  41.74 \\
		Basel CET 1 Ratio (adverse) &  10.32 &   7.07 &  10.54 &  12.18 &  33.96 \\
		Basel Leverage Ratio (2017) &   5.31 &   3.42 &   5.30 &   5.84 &  11.93 \\
		Basel Leverage Ratio (adverse) &   4.37 &   1.88 &   4.52 &   4.92 &  11.06 \\
		\bottomrule
	\end{tabular}
	%\tabnote{\textsuperscript{a}This footnote shows how to include footnotes to a table if required.}
	\label{Tb:CET}
\end{table}	

\subsection{Reconstruction of interbank network}

In order to test the reliability of contagious hierarchies identified by the proposed model, we employ 3 network reconstruction methods to build the asset/liability cross-holding network. As in \citet{FJ-Anand-vanLelyveld-Banai-Friedrich-Garratt-Halaj-Fique-Hansen-Jaramillo-Lee-MolinaBorboa-Nobili-Rajan-Salakhova-Silva-Silvestri-deSouza-2018-JFinancStab}, we call these 3 methods $\textit{Anan}$ \citep{FJ-Anand-Craig-VonPeter-2015-QuantFinanc}, $\textit{Ha\l{}a}$ \citep{FJ-Halaj-Kok-2013-ComputManageSci} and $\textit{Maxe}$ \citep{FJ-Upper-Worms-2004-EurEconRev}. Either of 3 methods can reconstruct interbank networks with aggregated assets and liabilities. However, the EBA dataset only provides asset exposures, no liability data. We refer to some empirical studies based on these data assuming that for bank $i$, the aggregated interbank assets  $\sum_{j} C_{i j} V_{j}$ equal to the aggregated interbank liabilities $\sum_{j} C_{j i} V_{i}$ \citep{FJ-Chen-Liu-Yao-2016-OperRes,FJ-Glasserman-Young-2015-JBankFinanc}.  We now give a brief description for these 3 methods. 

\subsubsection{Anan}
\cite{FJ-Anand-Craig-VonPeter-2015-QuantFinanc} propose a heuristic procedure for allocating links that combines elements from information theory and economic rationale. The authors argue that the Minimum Density (MD) method is suitable for sparse networks such as financial markets, and network is able to reconstructed by minimizing the cost of linkages. Its economic nature is similar with network design problems in transportation science.

Based on this method, $c$ is defined as the fixed cost of establishing a link, $N$ represents the number of banks. $C$ notes the matrix of aggregated exposure values. The aggregated interbank assets of bank $i$ are $a_{i}=\sum_{j=1}^{N}C_{ij}$, and its aggregated liabilities are $l_{i}=\sum_{j=1}^{N}C_{ji}$. Then, the MD method is formulated as:
\begin{equation}\label{Eq:anan}
	\begin{split}
		& \min c\sum_{i=1}^{N}\sum_{j=1}^{N}\bm{1}{ \{C_{ij}\ge 0\}}, ~~ s.t. \\
		& \sum_{j=1}^{N}C_{ij}=a_{i} ~~~ \forall i=1,2,...,N  \\
		& \sum_{i=1}^{N}C_{ij}=l_{j} ~~~ \forall j=1,2,...,N  \\
		& C_{ij}\geq 0 ~~~ \forall i,j,
	\end{split}
\end{equation}
where the integer function $\bm1$ is equal to one, only if bank $i$ lends to bank $j$, and zero otherwise. Here, the authors design a heuristic to solve this computationally expensive problem.

\subsubsection{Ha\l{}a}
\cite{FJ-Halaj-Kok-2013-ComputManageSci} propose a sampling method and each network is generated randomly based on the probability map. The size of linkage between banks is obtained by averaging many generated interbank structures. To generate one random network satisfied bilateral exposure requirements, an iterative algorithm is applied. At the initial network, assume that the possibility of all links is the same that all entries in the matrix $C^0$ are equal to zero, and the unmatched interbank assets and liabilities are initiated as $a^0=a $ and $l^0=l$. When iterating to the $k+1$ step, a pair of banks $(i,j)$ are randomly selected. Next, extract the random number $f$ from the unit interval to re-scale the matrix to update the weight $C^{k+1}_{ij}$ as follows: 
\begin{equation}\label{Eq:hala}
	C^{k+1}_{ij} = C^{k}_{ij} + f^{k+1}\min{\{a^k_i,l^k_j\}}
\end{equation}
and the unmatched assets and liabilities are:
\begin{equation}\label{Eq:hala2}
	a^{k+1}_{i} = a^{k}_{i} - \sum_{j=1}^{N}C^{k+1}_{ij} ~~and~~
	l^{k+1}_{j} = l^{k}_{j} - \sum_{i=1}^{N}C^{k+1}_{ij}
\end{equation}
The iteration is repeated until no more interbank assets are left to be assigned.

\subsubsection{Maxe}
$Maxe$ is the maximum entropy method, the basis of iterative methods \citep{FJ-Upper-Worms-2004-EurEconRev}. In the initial guess network, the exposure of bank $i$ to bank $j$ is equal to the aggregated interbank loans of bank $i$ multiplied by the aggregated interbank deposits of bank $j$, namely, $Q_{ij} = a_{i}l_{j}$. Next, the network is re-scaled until the constraints are satisfied. This entails maximizing the entropy function:
\begin{equation}\label{Eq:maxe}
	-\sum_{i,j}C_{i,j}\log(C_{i,j}/Q_{i,j}).
\end{equation}
According to \cite{FJ-Upper-Worms-2004-EurEconRev}, entropy optimization is a strict convex optimization problem, and yields a unique solution for the structure of interbank lending. It is solved numerically with the RAS algorithm that is commonly used in computing input-output tables \citep{FJ-Blien-1998-entropy}. \citet{FJ-Paltalidis-Gounopoulos-Kizys-Koutelidakis-2015-JBankFinanc} employ this method to reconstruct interbank network to study transmission channels of systemic risk. 

\subsubsection{Reconstructed European interbank networks}

Table~\ref{Tb:Network:Statistics} reports the network statistics we compute for reconstructed networks using above 3 approaches. It's shown that the reconstructed networks are very different. Network generated by $\textit{Maxe}$ has the largest number of links, the highest density and degree, so as to clustering and core size. This is because $\textit{Maxe}$ network is fully connected. Compared $\textit{Anan}$ with $\textit{Hala}$, we find that the $\textit{Anan}$ network is more sparse, having lower density and clustering, smaller average degree and core size. This is because $Anan$ is a minimum density method. The lender/borrower dependency is defined as the average of the market share of the largest borrower or lender, respectively. The HHI (Herfindahl-Hirschman Index) describes the concentration of both assets and liabilities. Due to the sparsity of $\textit{Anan}$ network, it's reasonable that this network has higher dependency and concentration. The assortativity characterizes the preference for a network's nodes to attach to others that are similar. Both $\textit{Anan}$ and $\textit{Hala}$ have negative assortativities, which is consistent with the statistic of the genuine interbank networks computed in \cite{FJ-Anand-vanLelyveld-Banai-Friedrich-Garratt-Halaj-Fique-Hansen-Jaramillo-Lee-MolinaBorboa-Nobili-Rajan-Salakhova-Silva-Silvestri-deSouza-2018-JFinancStab}. 

\begin{table}[!htb]
	\centering
	\caption{Network statistics for reconstructed interbank networks.}
	\smallskip
	\begin{tabular}{lccc}
			\toprule
			  & $\textit{Anan}$ & $\textit{Ha\l{}a}$ & $\textit{Maxe}$ \\
			  \midrule
			 Number of Links & 99 & 344 & 2256 \\
			Density &  4.388 & 15.248 & 100.000 \\
			Avg Degree &  2.063 &  7.167 & 47.000 \\
			Med Degree & 1 & 7 & 47 \\
			Assortativity & -0.308 & -0.321 &    NaN \\
			Clustering &  0.678 & 21.794 & 100.000 \\
			Lender Dependency & 83.718 & 57.910 & 10.708 \\
			Borrower Dependency & 86.334 & 71.905 & 10.708 \\
			Mean HHI Assets &  0.785 &  0.463 &  0.045 \\
			Median HHI Assets &  1.000 &  0.439 &  0.045 \\
			Mean HHI Liabilities &  0.824 &  0.639 &  0.045 \\
			Median HHI Liabilities &  1.000 &  0.620 &  0.045 \\
			Core Size (\% banks) & 10.417 & 18.750 & 97.917 \\
			\bottomrule
	\end{tabular}
	%\tabnote{\textsuperscript{a}This footnote shows how to include footnotes to a table if required.}
	\label{Tb:Network:Statistics}
\end{table}	

\begin{figure}[!htb]
	\centering
	\subfigure[Direct cross-holding $\mathbf{C}$ reconstructed by $\textit{Anan}$.]{%
		\resizebox*{7cm}{!}{\includegraphics{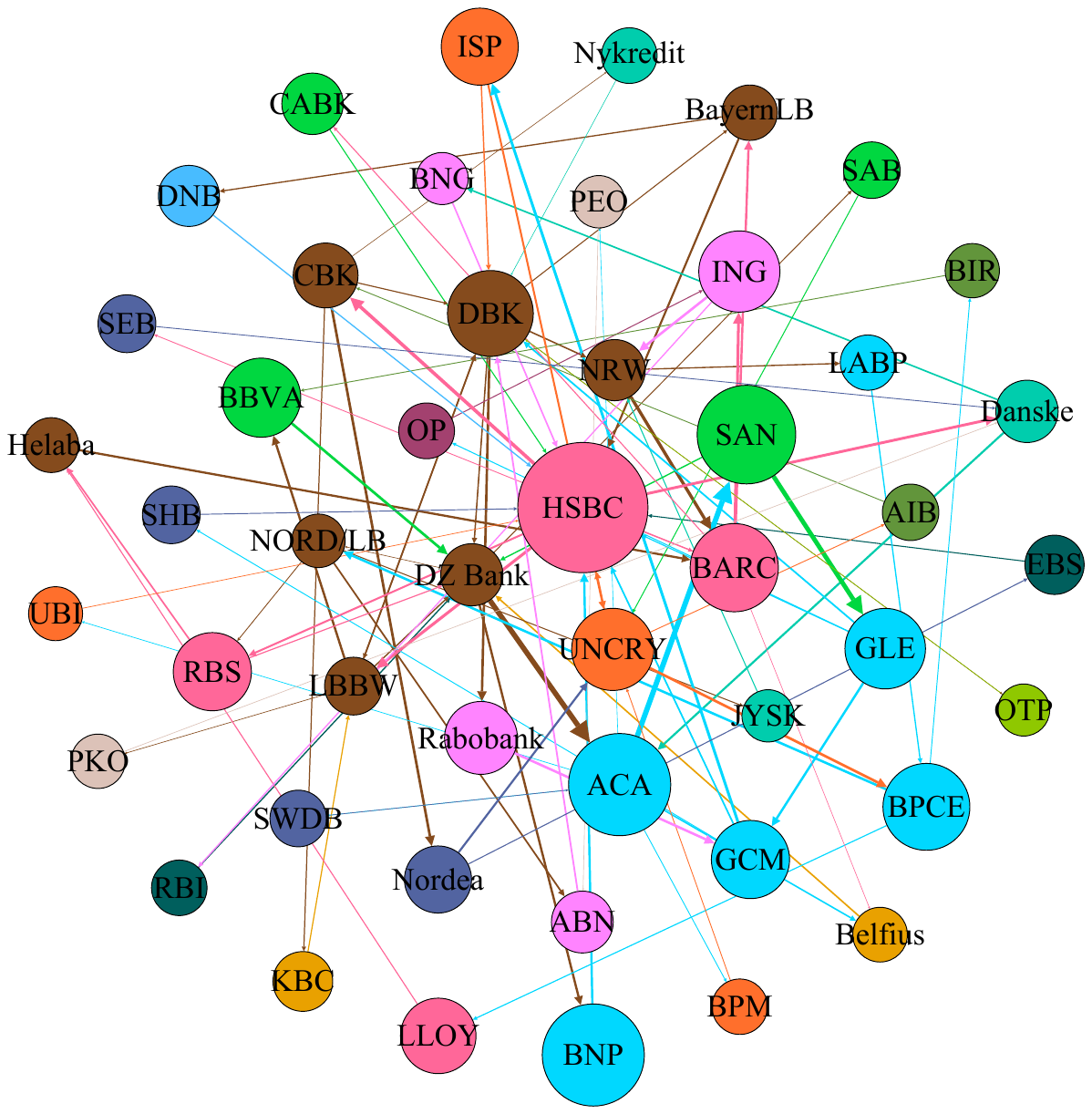}}}\hspace{5pt}
	\subfigure[Direct cross-holding $\mathbf{C}$ reconstructed by $\textit{Ha\l{}a}$.]{%
		\resizebox*{7cm}{!}{\includegraphics{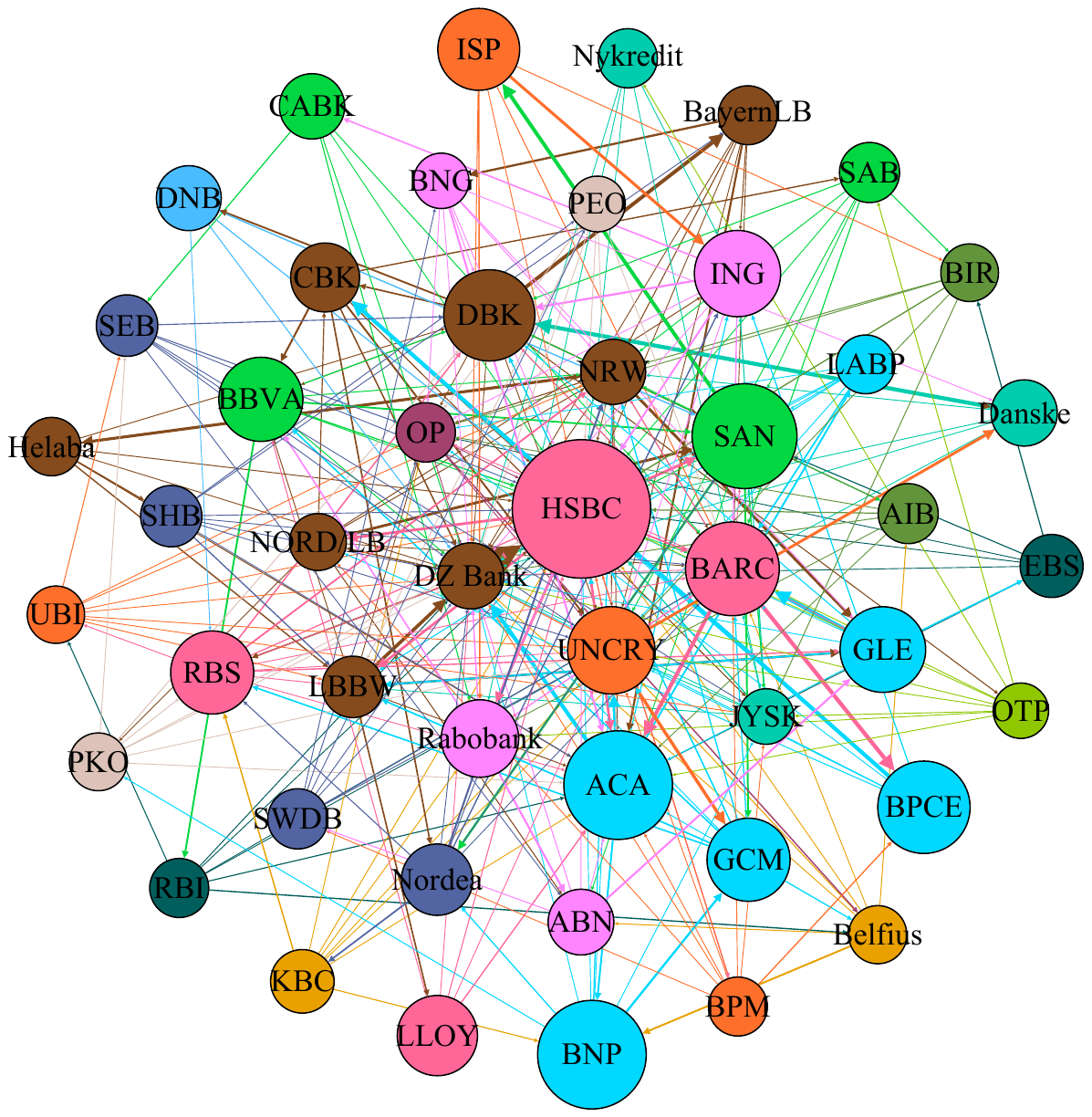}}}
	\caption{Direct cross-holding matrix $\mathbf{C}$ in European banking system reconstructed by $\textit{Anan}$ and $\textit{Ha\l{}a}$. The widths of the arrows are proportional to the sizes of the cross-holdings. The area of bank node is proportional to its equity value. The banks with the same color are belong to the same country.} \label{Fig:C}
\end{figure}

Figure~\ref{Fig:C} displays the European interbank metwork (direct cross-holding matrix $\mathbf{C}$) reconstructed by $\textit{Anan}$ and $\textit{Ha\l{}a}$ respectively. The widths of the arrows are proportional to the sizes of the cross-holdings. The area of bank node is proportional to its equity value. The banks with the same color are belong to the same country. The arrow direction means that the origin bank has claims on the destination bank. Consistent with Table~\ref{Tb:Network:Statistics}, the $\textit{Anan}$ network is more sparse than the $\textit{Ha\l{}a}$ network. We can also find that in both reconstructed networks, banks from the UK (the pink node), Germany (the brown node) and France (the blue node) are located in more central positions, showing that these banks are connected densely.

\begin{figure}[!htb]
	\centering
	\subfigure[Interdependent matrix $\mathbf{A}$ reconstructed by $\textit{Anan}$.]{%
		\resizebox*{7cm}{!}{\includegraphics{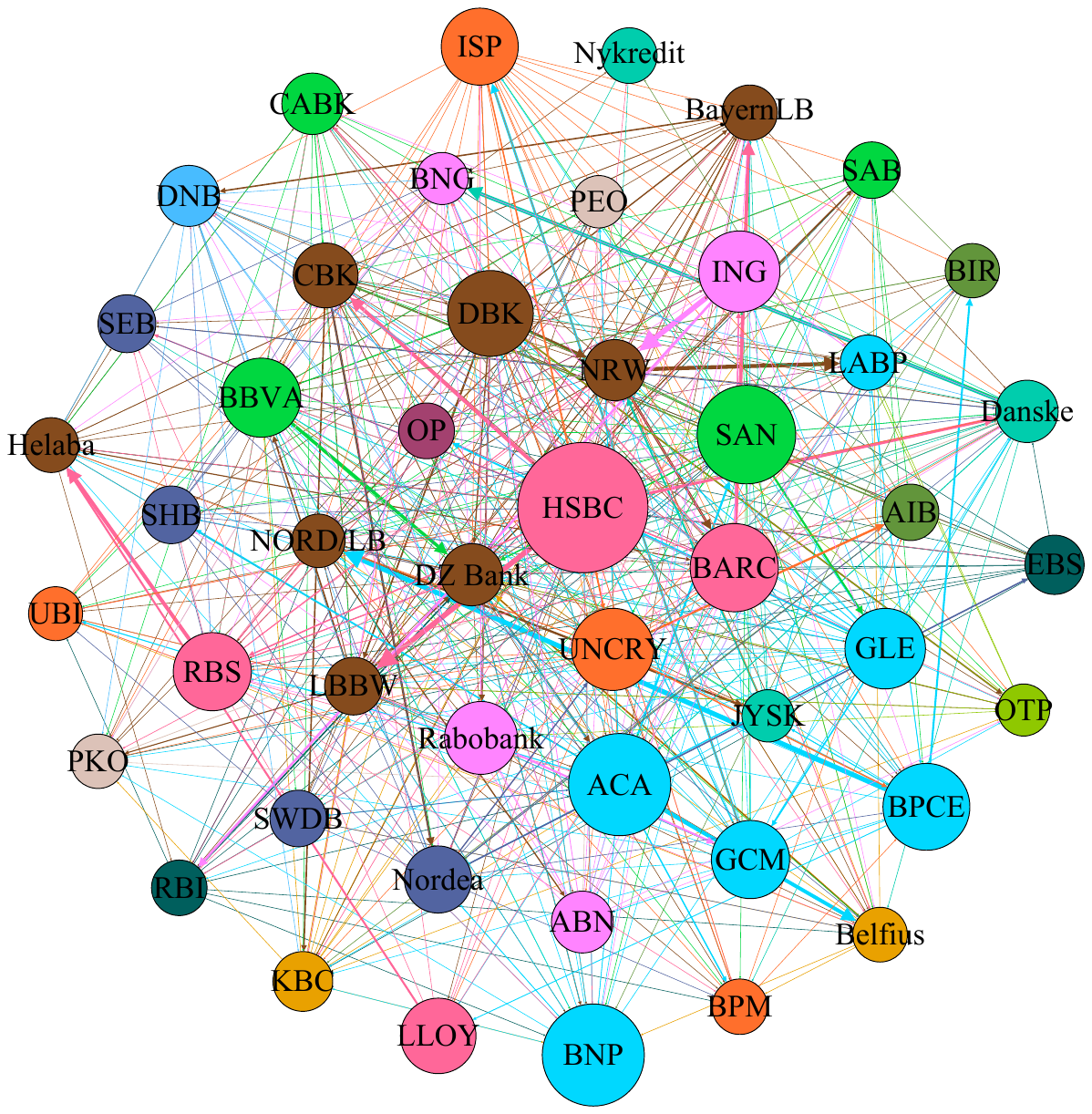}}}%\hspace{5pt}
	\subfigure[Interdependent matrix $\mathbf{A}$ reconstructed by $\textit{Ha\l{}a}$.]{%
		\resizebox*{7cm}{!}{\includegraphics{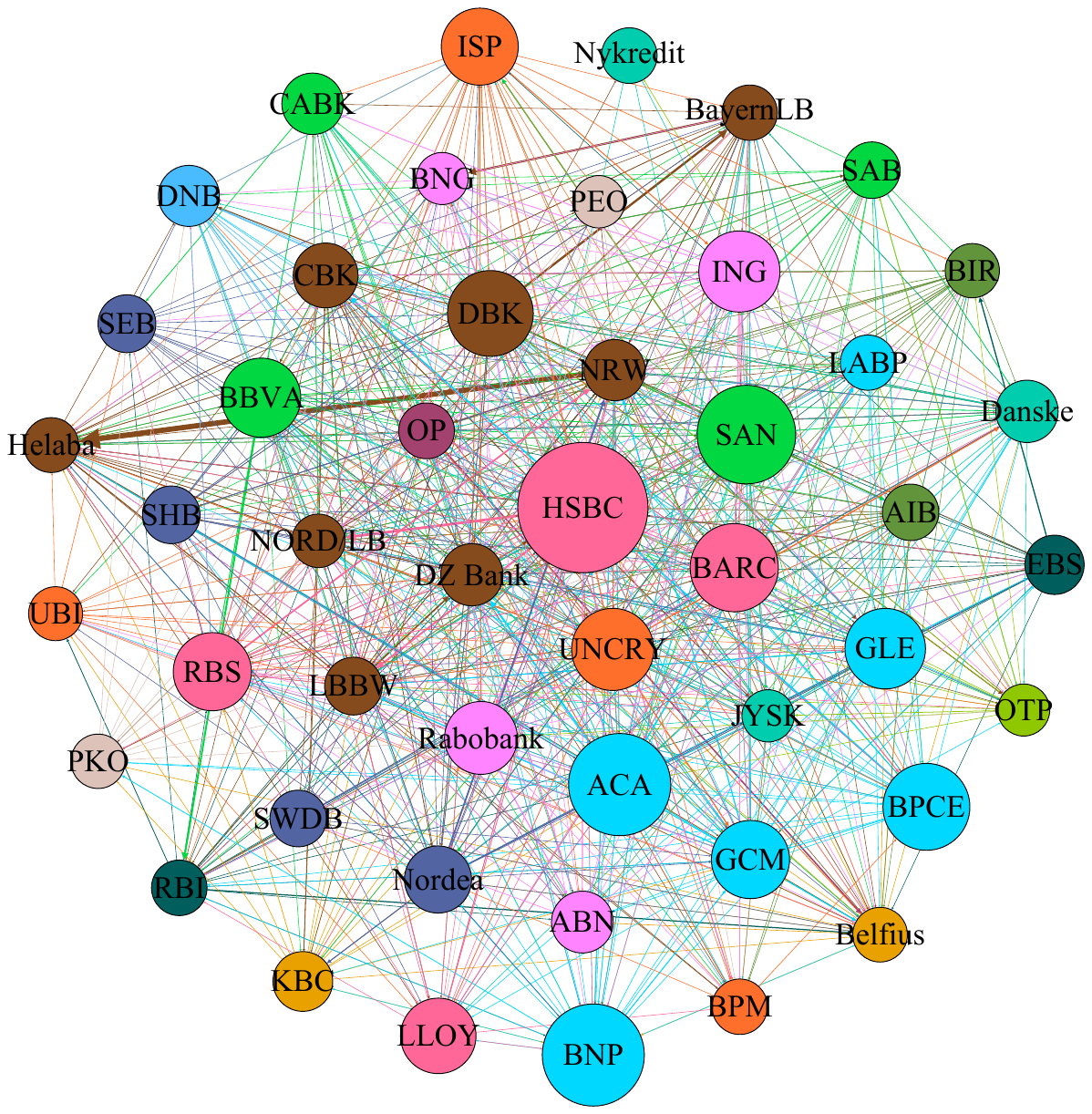}}}
	\caption{Interdependent matrix $\mathbf{A}$ in European banking system reconstructed by $\textit{Anan}$ and $\textit{Ha\l{}a}$. The widths of the arrows are proportional to the degrees of inter-dependency. The area of bank node is proportional to its equity value. The banks with the same color are belong to the same country.} \label{Fig:A}
\end{figure}

Figure~\ref{Fig:A} displays the interdependent matrix $\mathbf{A}$ in European banking system reconstructed by $\textit{Anan}$ and $\textit{Ha\l{}a}$. The widths of the arrows are proportional to the degrees of inter-dependency. Note that the interdependent matrix $\mathbf{A}$ not only describes the direct cross-holding among banks, but also the indirect claims on the external assets that other banks hold. Therefore, the interdependent network $\mathbf{A}$ are more dense than the direct interbank network $\mathbf{C}$. This is exactly explain what is interdependency and the difference between interdependency model and simple cross-holding model.

\subsection{Cascades}

\begin{table}[!htb]
	\centering
	\caption{Hierarchies of cascades in macroprudential stress test for the European banking system. Three reconstructure algorithms (i.e. $\textit{Anan}$,  $\textit{Ha\l{}a}$ and $\textit{Maxe}$) for the interbank cross-holding network are considered. This table reports the test results with different failure thresholds $\theta$ and different failure cost coefficients $\beta$. }
	\smallskip
	\begin{tabular}{lp{0.4\textwidth}cp{0.4\textwidth}}
			\toprule
			& \makecell[c]{$\beta$=0.3} && \makecell[c]{$\beta$=0.8}\\ \midrule
			\multicolumn{4}{l}{\textbf{Panel A:} $\textit{Anan}$, $\theta$=0.971} \\ \midrule
			First Failure & JYSK, GCM, Rabobank, DNB, SEB, SHB &&  JYSK, GCM, Rabobank, DNB, SEB, SHB\\
			\midrule
			\multicolumn{4}{l}{\textbf{Panel B:} $\textit{Anan}$, $\theta$=0.973} \\ \midrule
			First Failure & RBI, CBK, Danske, JYSK, BNP, ACA, GCM, HSBC, AIB, UBI, Rabobank, DNB, PEO, SEB, Nordea, SWDB, SHB && RBI, CBK, Danske, JYSK, BNP, ACA, GCM, HSBC, AIB, UBI, Rabobank, DNB, PEO, SEB, Nordea, SWDB, SHB \\
			Second Failure & DZ Bank,	BayernLB, ING && EBS, DZ Bank, BayernLB, UNCRY, ING \\
			Third Failure & LBBW,	BBVA &&  Belfius, LBBW, BBVA, OP \\
			Fourth Failure &                           && GLE \\
			\midrule
			\multicolumn{4}{l}{\textbf{Panel C:} $\textit{Ha\l{}a}$, $\theta$=0.971} \\ \midrule
			First Failure & JYSK, GCM, Rabobank, DNB, SEB, SHB &&  JYSK, GCM, Rabobank, DNB, SEB, SHB\\
			Second Failure & UBI &&  UBI\\
			Third Failure & RBI && RBI\\
			\midrule
			\multicolumn{4}{l}{\textbf{Panel D:} $\textit{Ha\l{}a}$, $\theta$=0.973} \\ \midrule
			First Failure & RBI, CBK, Danske, JYSK, BNP, ACA, GCM, HSBC, AIB, UBI, ING, Rabobank, DNB, PEO, SEB, Nordea, SWDB, SHB  && RBI, CBK, Danske, JYSK, BNP, ACA, GCM, HSBC, AIB, UBI, ING, Rabobank, DNB, PEO, SEB, Nordea, SWDB, SHB \\
			Second Failure & DZ Bank,	UNCRY && Belfius, DZ Bank, LBBW, BBVA, UNCRY \\
			Third Failure & LBBW   &&  OP \\
			\midrule
			\multicolumn{4}{l}{\textbf{Panel E:} $\textit{Maxe}$, $\theta$=0.971} \\ \midrule
			First Failure & JYSK, GCM, Rabobank, DNB, SEB, SHB &&  JYSK, GCM, Rabobank, DNB, SEB, SHB\\
			Second Failure &       &&  RBI \\
			\midrule
			\multicolumn{4}{l}{\textbf{Panel F:} $\textit{Maxe}$, $\theta$=0.973} \\ \midrule
			First Failure & RBI, CBK, Danske, JYSK, BNP, ACA, GCM, HSBC, AIB, UBI, ING, Rabobank, DNB, PEO, SEB, Nordea, SWDB, SHB  && RBI, CBK, Danske, JYSK, BNP, ACA, GCM, HSBC, AIB, UBI, ING, Rabobank, DNB, PEO, SEB, Nordea, SWDB, SHB \\
			Second Failure & LBBW &&  LBBW, BBVA, UNCRY \\
			Third Failure &     &&  DZ Bank, Helaba \\
			Fourth Failure &    &&  BayernLB, OP \\
			\bottomrule
	\end{tabular}
	%\tabnote{\textsuperscript{a}This footnote shows how to include footnotes to a table if required.}
	\label{Tb:Stress:Test}
\end{table}

To illustrate the hierarchical cascades, we consider the adverse scenario in EBA 2018 EU-wide stress test. The initial shock to the values of 20 types of external assets is extracted from the adverse scenario as of 2020. The failure thresholds $\underline{\mathbf{v}}$ are set to $\theta$ times the IFRS 9 restated figures at the end-2017 (which is the actual balance sheet data). Various levels of $\theta$ are chosen to test the cascade process. If a bank fails, then the loss in value is $\beta\underline{v}_i$, where $\beta$ is set to 0.3 for lower failure cost and 0.8 for higher failure cost.

We examine the results for $\textit{Anan}$ network, $\textit{Ha\l{}a}$ network and $\textit{Maxe}$ network respectively. In Table~\ref{Tb:Stress:Test}, Panel A and B display the hierarchies of cascades for $\textit{Anan}$ reconstructed network. In case of $\theta=0.971$, there are 5 banks hit its failure point under the initial shock. For both levels of failure costs, cascades do not occur. We then raise $\theta$ to 0.973 and see how cascades occur. In this case, there are 17 banks failed under the initial shock. Then DZ Bank, BayernLB and ING are triggered by a contagion when $\beta=0.3$. DZ Bank fails mainly due to its exposure to ACA and BNP. BayernLB fails mainly due to its exposure to HSBC and DNB. ING fails due to its exposure to RBI. When failure cost is raised to 0.8, two more banks (EBS and UNCRY) are failed in this hierarchy. In the next cascading round, when $\beta=0.3$, LBBW and BBVA are triggered to fail due to their exposures to the former two rounds of failed banks. For example, both LBBW and BBVA have claims on DZ Bank (see Figure~\ref{Fig:C}(a) and Figure~\ref{Fig:A}(a)). Pushing $\beta$ up to 0.8, there are two more banks (Belfius and OP) failed due to taking higher failure cost. In the final round, GLE also failed.

Panel C and D in Table~\ref{Tb:Stress:Test} display the hierarchies of cascades for $\textit{Ha\l{}a}$ reconstructed network. The initial failed banks are the same as the $\textit{Anan}$ cases.  However, cascades are triggered in case of $\theta=0.971$, that is, causing UBI to fail. This is due to the fact that cross-holding network reconstructed by $\textit{Ha\l{}a}$ has higher connection and density compared to the  $\textit{Anan}$ network. In the next round, UBI's failure further causes RBI to fail because RBI has claims on UBI (see Figure~\ref{Fig:C}(b) and Figure~\ref{Fig:A}(b)). When failure cost is raised to 0.8, there are no more banks failed. Pushing $\theta$ up to 0.973 leads to more banks failed and would cause failures at earlier levels, but would not change the ordering. Take $\beta=0.8$ for example, in case of $\theta=0.971$, the UBI failed at the second hierarchy, while in case of $\theta=0.973$, the UBI failed at the first hierarchy.

Panel E and F in Table~\ref{Tb:Stress:Test} display the hierarchies of cascades for $\textit{Maxe}$ reconstructed network. It is found that the cascading hierarchies are similar with the other two cases. It's reasonable since banks' external assets holdings weight more and play a key role in cascading dynamics. However, the structure of cross-holding network is also important for some specific banks. For example, Helaba failed in the case of  $\textit{Maxe}$, while not in the $\textit{Anan}$ and $\textit{Ha\l{}a}$ cases. Our results are consistent with \citet{FJ-Chen-Liu-Yao-2016-OperRes}, who find that the market liquidity effect has a greater potential than the network effect to cause systemic contagion. In addition, when $\beta=0.3$, the denser $\textit{Maxe}$ network leads to a weaker cascade effect than the other two networks do. This reflects that diversification plays a leading role when failure cost is lower.

\begin{table}[!htb]
	\centering
	\caption{Indirect loss for the European banking system under a stress test with failure threshold $\theta=0.973$ and failure cost coefficient $\beta=0.3$.}
	\smallskip
	\begin{tabular}{lrrr}
		\toprule
		& $\textit{Anan}$ & $\textit{Ha\l{}a}$ & $\textit{Maxe}$ \\
		\midrule
		First Failure (mln euro) & 165227.42 & 179893.33 & 180104.85 \\
		Second Failure (mln euro) & 24876.90 & 21678.20 & 3891.79 \\
		Third Failure (mln euro) & 19858.82 & 3887.53 &   0.00 \\
		Sum (mln euro) & 209963.14 & 205459.07 & 183996.64 \\
		Percentage (\% in total loss) &  23.57 &  23.18 &  21.27 \\
		Aggregate Vulnerability (\% in equity) &  15.57 &  15.24 &  13.65 \\
		\bottomrule
	\end{tabular}
	%\tabnote{\textsuperscript{a}This footnote shows how to include footnotes to a table if required.}
	\label{Tb:Indirect:Loss}
\end{table}	

Cascading failures cause additional losses in the banking system. Table~\ref{Tb:Indirect:Loss} reports these indirect losses in each round under a stress test with failure threshold $\theta=0.973$ and failure cost coefficient $\beta=0.3$. The ratio of indirect loss over total loss is, respectively, 23.57\% (209963.14 million euros) for $\textit{Anan}$ network, 23.18\% (205459.07 million euros) for $\textit{Ha\l{}a}$ network, and 21.27\% (183996.64 million euros) for $\textit{Maxe}$ network.  \cite{FJ-Greenwood-Landier-Thesmar-2015-JFinancEcon} propose an aggregate vulnerability measure, which calculates the percentage of aggregate bank equity that would be wiped out by deleveraging. This measure omits the direct losses, emphasizing only the spillover effect across the system. We translate this measure to our discontinuous setting. It measures the indirect equity losses wiped out by cascading failures. It is found that after the initial shock, bank equity is further wiped out by 15.57\% in $\textit{Anan}$, 15.24\% in $\textit{Ha\l{}a}$, and 13.65\% in $\textit{Maxe}$.  The diversified interbank structure in $\textit{Maxe}$ network reduces the cascading losses as we see in Table~\ref{Tb:Stress:Test}.

\subsection{Vulnerability rankings}

To quantitative the vulnerability for each bank, we take a test in the $\textit{Ha\l{}a}$ interbank network, with failure threshold $\theta=0.973$ and failure cost coefficient $\beta=0.3$. Table~\ref{Tb:Rank} lists the TOP 10 banks sorted according to total vulnerability. The vulnerability is assessed by summing the direct loss and indirect loss together. To see the impacts of two parts of losses, we also report direct vulnerability and indirect vulnerability. All measures are normalized by respective bank equity. The first 4 banks are so severely hit by both the initial shock and discontinuous failure cost that it wipes out their equity entirely. Overall, under this stress test, the initial adverse shock wipes out more than half equity, and the subsequent cascading defaults further wipe out about 30\% equity, causing these 10 banks most vulnerable.
 
 \begin{table}[!htbp]
 	\centering
 	\caption{TOP 10 banks sorted according to total vulnerability. It is tested in the $\textit{Ha\l{}a}$ interbank network, with failure threshold $\theta=0.973$ and failure cost coefficient $\beta=0.3$.}
 	\smallskip
 	\begin{tabular}{lllrrr}
 		\toprule
 		 Rank & Bank & Bank abbr. & Vulnerability & Direct  & Indirect \\
 		1 & JBayske Bank & JYSK & 100.00 &  76.14 &  23.86 \\
 		1 & Cooperatieve Rabobank U.A. & Rabobank & 100.00 &  63.54 &  36.46 \\
 		1 & Swedbank - group & SWDB & 100.00 &  68.15 &  31.85 \\
 		1 & Svenska Handelsbanken - group & SHB & 100.00 &  85.52 &  14.48 \\
 		5 & Unione di Banche Italiane Societa Per Azioni & UBI &  97.71 &  63.02 &  34.69 \\
 		6 & Skandinaviska Enskilda Banken - group & SEB &  95.23 &  63.50 &  31.73 \\
 		7 & Groupe Credit Mutuel & GCM &  93.86 &  55.55 &  38.31 \\
 		8 & Nordea Bank - group & Nordea &  93.05 &  58.86 &  34.19 \\
 		9 & Groupe Credit Agricole & ACA &  90.71 &  56.89 &  33.82 \\
 		10 & Danske Bank & Danske &  89.77 &  60.65 &  29.13 \\
 		\bottomrule
 	\end{tabular}
 %\tabnote{\textsuperscript{a}This footnote shows how to include footnotes to a table if required.}
\label{Tb:Rank}
\end{table}	

\section{Concluding remarks}
\label{S:Conclude}
In this paper, we derive the default cascade model and the fire-sale spillover model in a unified interdependent framework. We prove that these two loss mechanisms are essentially consistent, although one propagates losses in a discontinuous way and the other in a continuous way.

Based on an interdependent financial network, we have examined cascades in the European banking system. The interdependency means that the connections between banks include not only direct cross-holding (interbank network) but also indirect dependency by holding mutual assets outside the banking system (bipartite network).  Through analyzing bank's balance sheet, an equilibrium matrix is derived to characterize this interdependency. 

We use data extracted from the European Banking Authority to illustrate the interdependency. First, we collect 20 classes of  external assets mutually holding by 48 banks. For the cross-holding, interbank exposures are not available but the aggregated claims are public. Then we employ three network reconstruction methods to build the asset/liability cross-holding network. Finally, we compute the interdependency matrix. The interdependency network is much denser than the direct cross-holding network, showing the complex latent interaction among banks. 

Next we perform macroprudential stress tests for the European banking system, using the adverse scenario in EBA 2018
EU-wide stress test as the initial shock. For different reconstructed networks, we illustrate the hierarchical cascades and show that the failure hierarchies are roughly the same except for a few banks, reflecting the overlapping portfolio holding accounts for the majority of defaults. In addition, we calculate aggregate vulnerability for the banking system. We also display vulnerability rankings for individual banks. Our macroprudential stress test provides important information for bank's vulnerabilities. The results will form a solid ground for supervision and relevant management actions so as to strengthen their capital planning.

Clearly the above tests are based on moderate scenario taken by EBA (recalling that they assume GDP in the EU only decreases -1.2\%, -2.2\% and even increases 0.7\% as of 2018, 2019 and 2020 respectively), so that the default threshold must be set to a very high value (i.e. 0.97) to successfully trigger the initial failures. If the reverse scenario is more severe (e.g. the real state under COVID-19), the threshold value $\theta$ to trigger an initial failure can be set much smaller. In addition, the high sensitivity is also related with the high threshold. That is, if a small threshold (e.g. 0.5) can trigger an initial failure, the scope of optional thresholds could be much wider. For example, maybe we can choose 0.5, 0.6, 0.7 as threshold values to see the differences of cascade hierarchies. Nonetheless, we emphasize that understanding the interdependency network and the hierarchy of the cascades can help to improve policy intervention and implement rescue strategy.

%\section*{Acknowledgement(s)}
%
%We are grateful to the editor Iftekhar Hasan and two anonymous referees for their constructive comments and helpful suggestions.

\section*{Disclosure statement}

No potential conflict of interest was reported by the authors.

\bibliography{financial_contagion}

\appendix

\section{Proof of Proposition~\ref{Pro:existence}}
\label{S:Proof}

\begin{proof}[Proof]
	Observe that if $\mathbf{x} \leqslant \mathbf{y}$, then $\mathbf{b}(\mathbf{x}) \geqslant \mathbf{b}(\mathbf{y})$. Therefore, the map $\mathbf{v}\rightarrow - \mathbf{b}(\mathbf{v})$ is order-preserving, and hence so is the map $T$. In addition, for $\mathbf{v}\in [\mathbf{g}, \mathbf{v}_0]$, we have 
	\begin{equation}
		\mathbf{g}=\mathbf{A}(\mathbf{D}\mathbf{p}-\beta\underline{\mathbf{v}}) \leqslant T\mathbf{v}:=\mathbf{A}(\mathbf{D}\mathbf{p}-\mathbf{b}(\mathbf{v})) \leqslant \mathbf{A}\mathbf{D}\mathbf{p} = \mathbf{v}_0,
	\end{equation}
	so, $T$ is a self-map on $H:=[\mathbf{g}, \mathbf{v}_0]$ and $H$ is a complete lattice.  Following Tarski's fixpoint theorem \citep{FJ-Tarski-1955-PJM}, the map $T$ has a least fixed point $\mathbf{v_{*}}$ and a greatest fixed point $\mathbf{v^{*}}$ in $H$. 

As just discussed, the map $T$ is an order-preserving self-map on $[\mathbf{g}, \mathbf{v}_0]$, so $\{T^{k}\mathbf{g}\}$ is increasing. In addition, this sequence can take only finite values since $T$ has finite range. Let $\mathbf{v_{*}}$ be the limiting value and let $K$ be the number of iterations to attain $\mathbf{v_{*}}$. Then we have
\begin{equation}
	\mathbf{g} \leqslant T\mathbf{g} \leqslant T^2\mathbf{g} \leqslant \cdots \leqslant T^{K}\mathbf{g} = \mathbf{v_{*}},
\end{equation}
so $\mathbf{v_{*}}$ is a fixed point in $H$. Moreover, if $\mathbf{v^{\prime}}$ is any other fixed point of $T$ in $H$, then $\mathbf{g} < \mathbf{v^{\prime}}$ and hence $\mathbf{v_{*}}=T^{K}\mathbf{g} < T^{K}\mathbf{v^{\prime}} = \mathbf{v^{\prime}}$. Thus, $\mathbf{v_{*}}$ is the least fixed point of $T$.

Similar logic can be applied to prove that $\{T^{k}\mathbf{v}_0\}$ converges down to $\mathbf{v^{*}}$ in a finite number of iterations.
\end{proof}

\section{Bank list}
\setcounter{table}{0}
\begin{table}[!htbp]
	\centering
	\caption{Bank list.}
	\smallskip
	\begin{tabular}{llll}
		\toprule
		Country code      &   Country  &   Bank   &  Bank abbr. \\ \midrule
		AT	& Austria	& Raiffeisen Bank International AG  &RBI \\
		AT	&Austria	&Erste Group Bank AG	&EBS \\
		BE	&Belgium	&KBC Group NV	&KBC \\
		BE	&Belgium	&Belfius Banque SA	&Belfius \\
		DE	&Germany	&DZ BANK AG Deutsche Zentral-Genossenschaftsbank	&DZ Bank \\
		DE	&Germany	&Landesbank Baden-Wurttemberg	&LBBW \\
		DE	&Germany	&Deutsche Bank AG	&DBK \\
		DE	&Germany	&Commerzbank AG	&CBK \\
		DE	&Germany	&Norddeutsche Landesbank - Girozentrale -	&NORD/LB \\
		DE	&Germany	&Bayerische Landesbank	&BayernLB \\
		DE	&Germany	&Landesbank Hessen-Thuringen Girozentrale AdoR	&Helaba \\
		DE	&Germany	&NRW.BANK	&NRW \\
		DK	&Denmark	&Danske Bank	&Danske \\
		DK	&Denmark	&Jyske Bank	&JYSK \\
		DK	&Denmark	&Nykredit Realkredit	&Nykredit \\
		ES	 &Spain	         &Banco Santander S.A.	&SAN \\
		ES	 &Spain	        &Banco Bilbao Vizcaya Argentaria S.A.	&BBVA \\
		ES	 &Spain	        &CaixaBank, S.A.	&CABK \\
		ES	 &Spain	        &Banco de Sabadell S.A.	&SAB \\
		FI	  &Finland	     &OP Financial Group	&OP \\
		FR	 &France	   &BNP Paribas	&BNP \\
		FR	 &France	&Groupe Credit Agricole	&ACA \\
		FR	 &France	&Societe Generale S.A.	&GLE \\
		FR	 &France	&Groupe Credit Mutuel	&GCM \\
		FR	 &France	&Groupe BPCE	&BPCE \\
		FR	 &France	&La Banque Postale	&LABP \\
		GB	&United Kingdom	&Barclays Plc	&BARC \\
		GB	&United Kingdom	&Lloyds Banking Group Plc	&LLOY \\
		GB	&United Kingdom	&HSBC Holdings Plc	&HSBC \\
		GB	&United Kingdom	&The Royal Bank of Scotland Group Plc	&RBS \\
		HU	&Hungary	&OTP Bank Nyrt.	&OTP \\
		IE	  &Ireland	&Bank of Ireland Group  plc	&BIR \\
		IE	  &Ireland	&Allied Irish Banks Group plc	&AIB \\
		IT	  &Italy	&UniCredit S.p.A.	&UNCRY \\
		IT	  &Italy	&Intesa Sanpaolo S.p.A.	&ISP \\
		IT	  &Italy	&Banco BPM S.p.A.	&BPM \\
		IT	  &Italy	&Unione di Banche Italiane Societa Per Azioni	&UBI \\
		NL	&Netherlands	&N.V. Bank Nederlandse Gemeenten	&BNG \\
		NL	&Netherlands	&ABN AMRO Group N.V.	&ABN \\
		NL	&Netherlands	&ING Groep N.V.	&ING \\
		NL	&Netherlands	&Cooperatieve Rabobank U.A.	&Rabobank \\
		NO	&Norway	&DNB Bank Group	&DNB \\
		PL	 &Poland	&Powszechna Kasa Oszczednosci Bank Polski SA	&PKO \\
		PL	 &Poland	&Bank Polska Kasa Opieki SA	&PEO \\
		SE	 &Sweden	&Skandinaviska Enskilda Banken - group	&SEB \\
		SE	 &Sweden	&Nordea Bank - group	&Nordea \\
		SE	 &Sweden	&Swedbank - group	&SWDB \\
		SE	 &Sweden	&Svenska Handelsbanken - group	&SHB \\			
		\bottomrule
	\end{tabular}
	%\tabnote{\textsuperscript{a}This footnote shows how to include footnotes to a table if required.}
	\label{Tb:Bank:list}
\end{table}	

%\section{This is the title of the second appendix}

\end{document}